\documentclass[twocolumn,pra,showpacs,superscriptaddress]{revtex4}
\usepackage{amsbsy,bm}

\begin{document}

\title{Scattering theory for arbitrary potentials}

\author{A. S. Kadyrov}
\email{A.Kadyrov@murdoch.edu.au}
\affiliation{Centre for Atomic, Molecular and Surface Physics,
  Division of Science and Engineering, Murdoch University, Perth 6150,
  Australia}

\author{I. Bray}
\email{I.Bray@murdoch.edu.au}
\affiliation{Centre for Atomic, Molecular and Surface Physics,
  Division of Science and Engineering, Murdoch University, Perth 6150,
  Australia}

\author{A. M. Mukhamedzhanov}
\email{Akram@comp.tamu.edu}
\affiliation{Cyclotron Institute, Texas A\&M University, College
  Station, Texas 77843}

\author{A. T. Stelbovics}
\email{A.Stelbovics@murdoch.edu.au}
\affiliation{Centre for Atomic, Molecular and Surface Physics,
  Division of Science and Engineering, Murdoch University, Perth 6150,
  Australia}

\date{\today}



\begin{abstract}
  The fundamental quantities of potential scattering theory are
  generalized to accommodate long-range interactions.  New
  definitions for the scattering amplitude and wave operators valid
  for arbitrary interactions including potentials with a Coulomb tail
  are presented. It is shown that for the Coulomb potential the
  generalized amplitude gives the physical on-shell amplitude without
  recourse to a renormalization procedure.
\end{abstract}

\maketitle

\section{Introduction}


The two-body scattering problem is a central subject of quantum
mechanics. It is well known that conventional quantum collision theory
is formally valid only when the particles interact via short-range
potentials (see, e.g.~\cite{N82}).  In the time-dependent formulation,
formal scattering theory can be made to include Coulomb long-range
potentials by choosing appropriately modified time evolution
operators~\cite{D64,Dollard71}. This is equivalent to choosing various
forms of renormalization
methods~\cite{D68,V70Ru,GC74,Taylor75,Z76,H76} in the
time-independent formulation.

Though the renormalization theories lead to the correct cross sections
for the two-body problem, the results from these procedures cannot be
regarded as completely satisfactory. For instance, in screening-based
renormalization methods~\cite{D68,Taylor75} different ways of
shielding lead to different asymptotic forms for the scattering wave
function. Generally, these asymptotic forms differ from the physical
one obtained from the solution of the Schr\"odinger
equation~\cite{H85}. The weakest point about these methods, however,
is that they give rise to a scattering amplitude that does not exist
on the energy shell. This is because the amplitude obtained in these
methods has complex factors which are divergent on the energy
shell~\cite{OF60,S64,F64,NS70,D71,Taylor75}.  These factors, usually
containing branch point singularities, must be removed (renormalized)
before approaching the on-shell point.  Furthermore, the
renormalization factors depend on the way the limits are taken when
the on-shell point is approached.  Thus, the {\em ad-hoc}
renormalization procedure is based on prior knowledge of the exact
answer to compare with and has no {\em ab initio} theoretical
justification.  These issues are discussed in detail in the
comprehensive coverage of the subject given by \citet{H85}.

The motivation for the present work is to demonstrate that there is a
practical approach to the two-body collision problem with a
Coulomb-like potential that does not lead to the formal difficulties
described above.  Our approach is based on a representation of the
scattering amplitude written in a divergence-free surface-integral
form which is ideally suited for practical calculations. We build on a
recent formalism which has improved our understanding of the
three-body scattering processes~\citep{KMSB03l,KMSB04}. In this work
we consider local potentials. It is not difficult to show however that
the present results hold for nonlocal potentials as well.

In Section II we present a formal solution of the Schr\"odinger
equation which satisfies all necessary conditions imposed by the
long-range nature of the Coulomb interaction. We introduce new
well-defined forms of the scattering amplitude and wave operators for
two-body systems valid for arbitrary interactions. The new definitions
of the scattering amplitude not only cover arbitrary potentials but
also directly give the physical result. The relationship with
conventional formulations is discussed in Section III. Section IV
contains some concluding remarks including a brief discussion of the
utility of the two-body formalism for three-body Coulomb scattering
problems above the breakup threshold.

\section{Formalism for long-range potentials}

We consider scattering in a system of two particles 1 and 2
interacting via an arbitrary spherically-symmetric potential $V$
with the Coulomb long-range tail. Throughout the paper we use such
units that $\hbar=1$. A scattering state of this system is the
solution to the Schr\"odinger equation
\begin{eqnarray}
  (E -H) \psi^{\pm}_{\bm{k}} ( \bm{r}) =0 , \label{SE}
\end{eqnarray}
where $H=H_0+V$ is the total two-body Hamiltonian of the system, $H_0
=-\Delta_{\bm{r}}/2\mu $ is the free Hamiltonian, $E=k^2/2\mu$ is the
energy of the system, $\bm{r}$ is the relative coordinate of the
particles 1 and 2 and $\bm{k}$ is their relative momentum, $\mu$ is
their reduced mass. To be more specific we can assume that interaction
$V$ consists of some short-range part $V_s$ and the Coulomb potential
$V_c=z_{1} z_{2}/r$, where $z_1$ and $z_2$ are the charges of the
particles.

From all possible solutions to Eq.~(\ref{SE}) we should choose the
one satisfying the asymptotic boundary condition corresponding to
the physical scattering picture. When the potential has the
Coulomb tail the scattering wave function $\psi^{+}_{\bm{k}} (
\bm{r})$ asymptotically behaves, in the leading order, like the
Coulomb-modified plane wave and a Coulomb-modified outgoing
spherical wave
\begin{eqnarray}
  \psi^{+}_{\bm{k}} ( \bm{r}) &\stackrel{r \rightarrow
    \infty}{\sim}& e^{i\bm{k} \cdot \bm{r} +
    i{\gamma} \ln(k r - \bm{k} \cdot \bm{r})}
[1 +O(1/r)]
  \nonumber \\ &&
  + f(\widehat{\bm{k}} \cdot \widehat{\bm{r}})
  \frac{ e^{i {k} {r} -
      i{\gamma} \ln(2 k r)}} {r}
[1 +O(1/r)]
,
  \label{AB}
\end{eqnarray}
where $\gamma=z_{1} z_{2} \mu/k$ and $f$ is the scattering amplitude.
The second suitable solution $\psi^{-}_{\bm{k}} ( \bm{r})$
asymptotically behaves, in the leading order, like the
Coulomb-modified plane wave and a Coulomb-modified incoming spherical
wave
\begin{eqnarray}
  \psi^{-}_{\bm{k}} ( \bm{r}) &\stackrel{r \rightarrow
    \infty}{\sim}& e^{i\bm{k} \cdot \bm{r} -
    i{\gamma} \ln(k r + \bm{k} \cdot \bm{r})}
[1 +O(1/r)]
  \nonumber \\ &&
  + f^*(-\widehat{\bm{k}} \cdot \widehat{\bm{r}})
  \frac{ e^{-i {k} {r} +
      i{\gamma} \ln(2 k r)}} {r}
[1 +O(1/r)]
.
\nonumber \\
  \label{AB-}
\end{eqnarray}
Note that $\widehat{\bm{k}} \cdot \widehat{\bm{r}}\neq \pm 1$,
respectively for (\ref{AB}) and (\ref{AB-}). However, as we will see
below, in the asymptotic sense for which Eqs. (\ref{AB}) and
(\ref{AB-}) are written these forward/backward singularities have a
\mbox{$\delta$-function} nature and are, therefore, integrable. We
will also see that seemingly different forms of the Coulomb-modified
plane wave in Eqs.  (\ref{AB}) and (\ref{AB-}) asymptotically are
essentially the same function.

We can separate $\psi^{\pm}_{\bm{k}}$ into the so-called `incident'
and the `scattered' parts according to
\begin{eqnarray}
  \psi^{\pm}_{\bm{k}} ( \bm{r})&=&
  \phi^{\pm}_{\bm{k}} ( \bm{r})+
  \chi^{\pm}_{\bm{k}}( \bm{r}) , \label{split}
\end{eqnarray}
where $\phi^{\pm}_{\bm{k}}$ and $\chi^{\pm}_{\bm{k}}$ asymptotically
behave like the first and the second terms of Eqs.\ (\ref{AB}) and
(\ref{AB-}), respectively. The `unscattered' wave incident at
infinitely large distances is given by~\cite{H85}
\begin{eqnarray}
  \phi^{\pm}_{\bm{k}} ( \bm{r}) &=&
  e^{i\bm{k} \cdot \bm{r} \pm i \gamma \ln(k r \mp \bm{k} \cdot \bm{r})}
  \nonumber \\&&\times
  \sum_{n=0}^{\infty} [(\mp i \gamma)_n]^2
  \, (\mp i k r + i \bm{k} \cdot \bm{r})^{-n}/n!
  , \label{casseries}
\end{eqnarray}
where $(z)_n=z(z+1) \cdots (z+n-1)$. As to the unknown scattered wave
$\chi^{\pm}_{\bm{k}}$, the form of Eqs.~(\ref{AB}) and (\ref{AB-})
suggest that its leading order term in the asymptotic region already
contains all the scattering information we want. The next order terms
simply repeat this information. Therefore, all we need for extracting
the scattering amplitude is the leading-order asymptotic term of the
scattered wave~$\chi^{\pm}_{\bm{k}}$.

Let us denote $\phi^{(0)\pm}_{\bm{k}} ( \bm{r})$ the first term of the
incident wave $\phi^{\pm}_{\bm{k}}$,
\begin{eqnarray}
  \phi^{(0)\pm}_{\bm{k}} ( \bm{r}) &
=& e^{i\bm{k} \cdot \bm{r} \pm i{\gamma} \ln(k r \mp \bm{k}
\cdot \bm{r})}
, \label{cas}
\end{eqnarray}
and by $\phi^{(1)\pm}_{\bm{k}} ( \bm{r})$ the second term etc.:
\begin{eqnarray}
  \phi^{\pm}_{\bm{k}} ( \bm{r}) &
  =& \phi^{(0)\pm}_{\bm{k}} ( \bm{r}) + \phi^{(1)\pm}_{\bm{k}} ( \bm{r}) + \phi^{(2)\pm}_{\bm{k}} ( \bm{r}) + \cdots .
\label{incidentwave}
\end{eqnarray}
Then Eq.~(\ref{SE}) can be written in the form
\begin{eqnarray}
(E -H) \left[ \psi^{\pm}_{\bm{k}} ( \bm{r})
-\phi^{(0)\pm}_{\bm{k}} ( \bm{r})  \right] &=& (H -E)
\phi^{(0)\pm}_{\bm{k}} ( \bm{r}) .
\label{SE2}
\end{eqnarray}
If we introduce Green's function according to
\begin{eqnarray}
(E -H) G(\bm{r},\bm{r}',E) =\delta(\bm{r}-\bm{r}') , \label{GF}
\end{eqnarray}
and apply it onto both sides of Eq.\ (\ref{SE2}) we have
\begin{eqnarray}
\int d\bm{r}'
G(
\bm{r}, \bm{r}';
E \pm i\epsilon)
(E-\roarrow{H}) \left[ \psi^{\pm}_{\bm{k}} ( \bm{r}')
-\phi^{(0)\pm}_{\bm{k}} ( \bm{r}')  \right]
\nonumber \\
=
\int d\bm{r}'
G(
\bm{r}, \bm{r}';
E \pm i\epsilon)
(\roarrow{H}-E) \phi^{(0)\pm}_{\bm{k}} ( \bm{r}') ,
\label{neweq1}
\end{eqnarray}
where $\epsilon$ is a small positive parameter and limit as $\epsilon
\to 0$ is assumed~\cite{endnote}.
We used an arrow on the differential Hamiltonian operator to show the
direction in which it acts. We emphasize here a subtle point that the
operator $G(\bm{r},\bm{r}',E) (E -\roarrow{H})$ in Eq.~(\ref{neweq1})
does not act like a \mbox{$\delta$-function}. In other words, though
\begin{eqnarray}
G(\bm{r},\bm{r}',E) (E -\loarrow{H})=(E -\roarrow{H})
G(\bm{r},\bm{r}',E) \equiv \delta(\bm{r}-\bm{r}'),
\nonumber \\
\end{eqnarray}
however, in general
\begin{eqnarray}
G(\bm{r},\bm{r}',E) (E -\roarrow{H}) \neq \delta(\bm{r}-\bm{r}').
\end{eqnarray}
The reason is that the operator $G(\bm{r},\bm{r}',E) (E -\roarrow{H})$
produces an integral that has a surface-integral component. In order
for the action of the operator $G(\bm{r},\bm{r}',E) (E -\roarrow{H})$
to be equal to that of $G(\bm{r},\bm{r}',E) (E -\loarrow{H})$ this
surface-integral component should be zero.  A similar problem has been
discussed by \citet{Gloeckle83} in relation to the three-body
Lippmann-Schwinger equations. Using an operator similar to ours but
with the Hamiltonian $H$ and the Green's function $G$ replaced by some
channel Hamiltonian and corresponding Green's function Gl\"ockle
showed that surface integrals of the same nature only disappear if the
operator acts on a function which vanishes sufficiently quickly at
infinity. It is not difficult to demonstrate that the same conclusion
also applies to operator $G(\bm{r},\bm{r}',E) (E -\roarrow{H})$. Since
the wave functions $\psi^{\pm}_{\bm{k}} ( \bm{r})
-\phi^{(0)\pm}_{\bm{k}} ( \bm{r})$ and $\phi^{(0)\pm}_{\bm{k}} (
\bm{r})$ are examples of functions which do not vanish at infinity the
surface integral terms generated do not vanish.

On the right hand side of Eq.\ (\ref{neweq1}) we have a purely
scattered wave generated from $\phi^{(0)\pm}_{\bm{k}} ( \bm{r})$.
Therefore, we denote the result of the action of the integral
operator on the left hand side of Eq.\ (\ref{neweq1}) as
\begin{eqnarray}
\int d\bm{r}'
G(
\bm{r}, \bm{r}';
E \pm i\epsilon)
(E-\roarrow{H}) \left[ \psi^{\pm}_{\bm{k}} ( \bm{r})
-\phi^{(0)\pm}_{\bm{k}} ( \bm{r})  \right]
\nonumber \\
\equiv \chi^{(0)\pm}_{\bm{k}}
(\bm{r}),
\end{eqnarray}
which is a part of the scattered wave
$\chi^{\pm}_{\bm{k}} (\bm{r})$.
Then for the part of the total scattering wave
function $\psi^{(0)\pm}_{\bm{k}}$ developed from the leading term
of the incident wave,
\begin{eqnarray}
\psi^{(0)\pm}_{\bm{k}}
\equiv
\phi^{(0)\pm}_{\bm{k}} ( \bm{r}) + \chi^{(0)\pm}_{\bm{k}}
(\bm{r}) ,
\end{eqnarray}
we can write
\begin{eqnarray}
\psi^{(0)\pm}_{\bm{k}} ( \bm{r}) &=& \phi^{(0)\pm}_{\bm{k}} (
\bm{r}) +
\int d\bm{r}'
G(\bm{r}, \bm{r}';E \pm i\epsilon)
\nonumber \\&& \times
(\roarrow{H}-E) \phi^{(0)\pm}_{\bm{k}} ( \bm{r}') .
\label{formalsol0}
\end{eqnarray}
Note that although $\phi^{(0)\pm}_{\bm{k}} ( \bm{r})$ is just the
leading order term of the incident wave, $\psi^{(0)\pm}_{\bm{k}}
(\bm{r})$ includes scattered waves of all orders through the
Green's function. Moreover, as we will see below the scattered
wave in Eq.~(\ref{formalsol0}) is in fact the only scattered part
of the total wave function $\psi^{\pm}_{\bm{k}} (\bm{r})$.

Similarly, if $\phi^{(1)\pm}_{\bm{k}}$ is the next-to-the-leading
order term of the incident wave then
\begin{eqnarray}
(E -H) \left[ \psi^{\pm}_{\bm{k}} ( \bm{r})
-\phi^{(1)\pm}_{\bm{k}} ( \bm{r})  \right] &=& (H -E)
\phi^{(1)\pm}_{\bm{k}} ( \bm{r}) .
\nonumber \\
\label{SE21}
\end{eqnarray}
Therefore, for the part of the scattering wave $\psi^{\pm}_{\bm{k}}$
developed from $\phi^{(1)\pm}_{\bm{k}}$ term [i.e., for
$\psi^{(1)\pm}_{\bm{k}} \equiv \phi^{(1)\pm}_{\bm{k}} ( \bm{r}) +
\chi^{(1)\pm}_{\bm{k}} (\bm{r})$] we have
\begin{eqnarray}
\psi^{(1)\pm}_{\bm{k}} ( \bm{r}) &=& \phi^{(1)\pm}_{\bm{k}} (
\bm{r}) +
\int d\bm{r}'
G(\bm{r}, \bm{r}';E \pm i\epsilon) \nonumber \\&& \times
(\roarrow{H}-E) \phi^{(1)\pm}_{\bm{k}} ( \bm{r}') .
\label{formalsol1}
\end{eqnarray}
By direct substitution of Eqs.~(\ref{formalsol0}) and
(\ref{formalsol1}) into Eq.~(\ref{SE}) and using Eq.~(\ref{GF}) we
verify that
$\psi^{(0)\pm}_{\bm{k}} ( \bm{r})$ and $\psi^{(1)\pm}_{\bm{k}} (
\bm{r})$ satisfy Eq.~(\ref{SE}).  Obviously, continuing this
procedure $\psi^{\pm}_{\bm{k}} ( \bm{r})$ can be formally
reconstructed:
\begin{eqnarray}
  \label{formalsolful}
  \psi^{\pm}_{\bm{k}} ( \bm{r}) &=& \psi^{(0)\pm}_{\bm{k}} ( \bm{r})
  +  \psi^{(1)\pm}_{\bm{k}} ( \bm{r}) + \psi^{(2)\pm}_{\bm{k}} ( \bm{r}) + \dots
\end{eqnarray}
The latter can simply be written as
\begin{eqnarray}
  \label{formalsolfulshort}
  \psi^{\pm}_{\bm{k}} ( \bm{r}) &=& \phi^{\pm}_{\bm{k}} ( \bm{r})
  +  \int d\bm{r}'
G(\bm{r}, \bm{r}';E \pm i\epsilon)
(\roarrow{H}-E) \phi^{\pm}_{\bm{k}} ( \bm{r}') .
\nonumber \\
\end{eqnarray}
It is easily verified (by substitution) that this is a formal
solution to Eq.\ (\ref{SE}). Generally, this solution is not
unique. The physical solution is the one which satisfies the
boundary conditions specified in Eq.~(\ref{AB}) or~(\ref{AB-}). At
first sight the integral term in Eq.~(\ref{formalsolfulshort})
seems to involve a nonintegrable singularity at the origin for the
higher-order terms of the incident wave. However, as we will now
demonstrate, this is not the case.

We begin by noting as mentioned earlier that the leading--order
asymptotic term of $\psi^{\pm}_{\bm{k}} ( \bm{r})$ already contains
all the scattering information we need, provided
Eq.~(\ref{formalsolfulshort}) is the solution which has the required
asymptotic behavior. In other words, for the purpose of extracting the
scattering amplitude we have to verify that Eq.\
(\ref{formalsolfulshort}) asymptotically behaves
as Eqs.~(\ref{AB}) or~(\ref{AB-}). To this end let us write the
full set of eigenstates of Hamiltonian $H$ as
$\{\psi^{\pm}_{\bm{k}},\varphi_{n} \}$, where $\varphi_{n} $ are
the eigenfunctions corresponding to negative discrete
eigenvalues~$E_n$. Then we express the Green's function in its
spectral decomposition as
\begin{eqnarray}\label{spectr}
G(\bm{r}, \bm{r}';E \pm i\epsilon) &=& \int \frac{d {\bm k}'}{(2
\pi)^3} \frac{\psi^{\mp}_{\bm{k}'} ( \bm{r})\psi^{\mp*}_{\bm{k}'}(
\bm{r}')}{E-k'^2/2\mu\pm i\epsilon} \nonumber \\&& + \sum_n
\frac{\varphi_{n} ( \bm{r}) \varphi^{*}_{n}( \bm{r}')}{E-E_n \pm
i\epsilon} .
\end{eqnarray}
Using this we can write Eq.~(\ref{formalsolfulshort}) as
\begin{eqnarray}
\psi^{+}_{\bm{k}} ( \bm{r}) &=& \phi^{+}_{\bm{k}} ( \bm{r})
\nonumber \\&&
+
\int \frac{d {\bm k}'}{(2 \pi)^3}
\frac{
\big<\psi^{-}_{\bm{k}'} |\roarrow{H}-E|
\phi^{+}_{\bm{k}}\big>
\, \psi^{-}_{\bm{k}'} ( \bm{r})}
{E-k'^2/2\mu+i\epsilon}
+ \cdots,
\nonumber \\
\label{formalsolfull2bra}
\end{eqnarray}
where dots indicate the contribution from the bound states. As we
are interested only in the asymptotic behavior 
when ${r} \to \infty$ 
the bound states do not contribute.

Consider now
\begin{eqnarray}
  \big<\psi^{-}_{\bm{k}'} |\roarrow{H}-E|
  \phi^{+}_{\bm{k}}\big> &\equiv&
  \int d\bm{r}
  \psi^{-*}_{\bm{k}'} ( \bm{r})
  (\roarrow{H}-E) \phi^{+}_{\bm{k}} ( \bm{r})
  ,
  \nonumber \\
\end{eqnarray}
the integral entering Eq.~(\ref{formalsolfull2bra}). This is a kind
of integral with a possible nonvanishing surface-integral component we
mentioned earlier.  Using Eq.~(\ref{SE}) we can write
\begin{eqnarray}
\big<\psi^{-}_{\bm{k}'} |\roarrow{H}-E|
\phi^{+}_{\bm{k}}\big>
&=& \big<\psi^{-}_{\bm{k}'}
|-\loarrow{H}+E+\roarrow{H}-E| \phi^{+}_{\bm{k}}\big>
\nonumber \\
&=& -\big<\psi^{-}_{\bm{k}'}
|\loarrow{H}_0-\roarrow{H}_0| \phi^{+}_{\bm{k}}\big>
.
\end{eqnarray}
Using Green's theorem this volume-integral can be transformed to a
surface integral.
 We have
\begin{eqnarray}
\big<\psi^{-}_{\bm{k}'} |\roarrow{H}-E|
\phi^{+}_{\bm{k}}\big>
&=& - \frac{1}{2\mu} \lim_{r \rightarrow
\infty} r^2 \int \hat{\bm{r}} d\hat{\bm{r}} \nonumber \\&& \times \left[
 \psi^{-*}_{\bm{k}'} \nabla_{\bm{r}} \phi^{+}_{\bm{k}}
- \phi^{+}_{\bm{k}} \nabla_{\bm{r}} \psi^{-*}_{\bm{k}'}
\right] 
\nonumber \\
&=& - \frac{1}{2\mu} \lim_{r \rightarrow
\infty} r^2 \int d\hat{\bm{r}} \nonumber \\&& \times \left[
 \psi^{-*}_{\bm{k}'} \frac{\partial \phi^{+}_{\bm{k}} }{\partial r}
- \phi^{+}_{\bm{k}} \frac{\partial \psi^{-*}_{\bm{k}'} }{\partial r}
\right] .
\end{eqnarray}
Thus in the surface-integral form our integral depends only on
asymptotic behavior of the participating functions. Now noting
that
\begin{eqnarray}
  \phi^{\pm}_{\bm{k}} ( \bm{r}) &\stackrel{r \rightarrow
    \infty}{\sim}
  & \phi^{(0)\pm}_{\bm{k}} ( \bm{r})
[1 +O(1/r)]
, \label{incidentwaveas}
\end{eqnarray}
it is easy to see that
\begin{eqnarray}
\big<\psi^{-}_{\bm{k}'} |\roarrow{H}-E|
\phi^{+}_{\bm{k}}\big>
&=& - \big<\psi^{-}_{\bm{k}'} |\loarrow{H}_0-\roarrow{H}_0|
\phi^{(0)+}_{\bm{k}}\big>
\nonumber \\
&=& \big<\psi^{-}_{\bm{k}'} |\roarrow{H}-E|
\phi^{(0)+}_{\bm{k}}\big>
.
\label{newres}
\end{eqnarray}
In combination with Eq.~(\ref{formalsolfull2bra}) this result
in fact means that
\begin{eqnarray}
  \label{kernel}
\lefteqn{
\int d\bm{r}'
G(\bm{r}, \bm{r}';E \pm i\epsilon)
(\roarrow{H}-E) \phi^{\pm}_{\bm{k}} ( \bm{r}')}    \nonumber \\&&
=
\int d\bm{r}'
G(\bm{r}, \bm{r}';E \pm i\epsilon)
(\roarrow{H}-E) \phi^{(0)\pm}_{\bm{k}} ( \bm{r}') ,
\end{eqnarray}
i.e.\, all the scattered wave is generated from the first term of
the incident wave $\phi^{(0)+}_{\bm{k}}$. In other words, somewhat
unexpectedly, no scattered wave is produced from the remaining
part of $\phi^{+}_{\bm{k}}$. This also shows that
Eq.~(\ref{formalsolfulshort}) does not contain nonintegrable
singularities for small $\bm{r}'$.

Thus, using Eq.~(\ref{newres}) we can write
Eq.~(\ref{formalsolfull2bra}) as
\begin{eqnarray}
\psi^{+}_{\bm{k}} ( \bm{r}) &=& \phi^{+}_{\bm{k}} ( \bm{r})
+
\int \frac{d {\bm k}'}{(2 \pi)^3}
\frac{
T(\bm{k}',\bm{k})
\, \psi^{-}_{\bm{k}'} ( \bm{r})}
{E-k'^2/2\mu+i\epsilon}
+ \cdots,
\nonumber \\
\label{formalsolfull2}
\end{eqnarray}
where dots still indicate only the contribution from the bound
states. In the above equation we have introduced the notation
\begin{eqnarray}
T(\bm{k}',\bm{k}) &=& \big<\psi^{-}_{\bm{k}'} |\roarrow{H}-E|
\phi^{(0)+}_{\bm{k}}\big> , \label{prior1}
\end{eqnarray}
anticipating that $T(\bm{k}',\bm{k})$ is the desired scattering
\mbox{$T$--matrix}.
However this requires justification that we now
provide. In order to prove this let us first expand the scattering
wave function $\psi^{+}_{\bm{k}} ( \bm{r})$ according to
\begin{eqnarray}
\psi^{+}_{\bm{k}} ( \bm{r})&=&
\sum_{l,m} i^l e^{i\sigma_l(k)} \chi_{l}(k_{},r_{})
Y^*_{l,m}(\widehat{\bm{k}}_{}) Y_{l,m}(\widehat{\bm{r}}_{}),
\label{psi+inpw}
\end{eqnarray}
where
$\sigma_l(k)$
is the (total)
phase shift. The radial
functions asymptotically behave according to
\begin{eqnarray}
\chi_{l}(k,r) &\stackrel{r \rightarrow \infty}{\sim} & \frac{4
\pi}{k r} \sin\left[ { k r - \gamma \ln |2kr| - l\pi/2+\sigma_l(k)
} \right]
. \nonumber \\
\label{varphias}
\end{eqnarray}
Therefore, substituting Eq. (\ref{varphias}) into (\ref{psi+inpw}) we
get, after some 
algebra,
\begin{eqnarray}
\psi^{+}_{\bm{k}} ( \bm{r})
&\stackrel{r \rightarrow
\infty}{\sim} &
\frac{2 \pi}{i k r} \left[ e^{ ik r -i\gamma \ln |2kr| }
\delta(\widehat{\bm{k}}-\widehat{\bm{r}}) \right. \nonumber \\ &&
\left. -e^{ -ik r +i\gamma \ln |2kr| }
\delta(\widehat{\bm{k}}+\widehat{\bm{r}}) \right] \nonumber \\ &&
+\frac{e^{ ik r -i\gamma \ln |2kr| }}{r}
\frac{2 \pi}{i k }
\sum_{l,m} (e^{2i\sigma_l(k)}-1)
\nonumber \\ && \times
Y^*_{l,m}(\widehat{\bm{k}}_{}) Y_{l,m}(\widehat{\bm{r}}_{})
.
\label{phi+as2}
\end{eqnarray}
Since~\cite{endnote2}
\begin{eqnarray}
\frac{2 \pi}{i k } \sum_{l,m} (e^{2i\sigma_l(k)}-1)
Y^*_{l,m}(\widehat{\bm{k}}_{}) Y_{l,m}(\widehat{\bm{r}}_{}) &=&
f(\widehat{\bm{k}} \cdot \widehat{\bm{r}}) , \label{finpw}
\end{eqnarray}
we may conclude that Eq. (\ref{phi+as2}) is equivalent to Eq.
(\ref{AB}). The amplitude $f$ is still unknown, however, from here
we get an interesting result that
the first term of Eq. (\ref{phi+as2}) is an asymptotic form of the
distorted plane wave $\phi^{(0)+}_{\bm{k}}( \bm{r})$. Repeating
the same reasoning for $\psi^{-}_{\bm{k}}( \bm{r}) $ [or using
$\psi^{-}_{\bm{k}} ( \bm{r}) = \psi^{+*}_{-\bm{k}} ( \bm{r})$]
we establish that exactly the same term
is the asymptotic form of $ \phi^{(0)-}_{\bm{k}}( \bm{r}) $ as
well. That is,
\begin{eqnarray}
e^{i\bm{k} \cdot \bm{r} \pm i\gamma \ln(k r \mp \bm{k} \cdot
\bm{r})} &\stackrel{r \rightarrow \infty}{\sim}&  \frac{2 \pi}{i k
r} \left[ e^{ ik r -i\gamma \ln |2kr| }
\delta(\widehat{\bm{k}}-\widehat{\bm{r}}) \right. \nonumber \\ &&
\left. -e^{ -ik r +i\gamma \ln |2kr| }
\delta(\widehat{\bm{k}}+\widehat{\bm{r}}) \right]
. \nonumber \\
\label{asdpwgen}
\end{eqnarray}
In the absence of the long-range distortion Eq.\ (\ref{asdpwgen})
transforms to the familiar asymptotic form of the plane wave (see,
e.g., \cite{M65}):
\begin{eqnarray}
e^{i\bm{k} \cdot \bm{r} } &\stackrel{r \rightarrow
\infty}{\sim}&  \frac{2 \pi}{i k r} \left[ e^{ ik r } \delta(\widehat{\bm{k}}-\widehat{\bm{r}})
-e^{ -ik r } \delta(\widehat{\bm{k}}+\widehat{\bm{r}})
\right]
.
\label{aspw}
\end{eqnarray}
Thus the beauty of the result is that the notorious
forward/backward singularity of the distorted plane waves
$\phi^{(0)+}_{\bm{k}} (\bm{r})$/$\phi^{(0)-}_{\bm{k}}( \bm{r})$ is
shown to be no more problematic than a 
\mbox{$\delta$-function}. Therefore, in the asymptotic region the
distorted plane wave can be treated much like the plane wave. In
this regard it is worth of mentioning that the Coulomb Green's
function in momentum space has also been shown to have distorted
pole singularities~\cite{V72Ru} which follow from the
existence of the Dollard wave operators.

Returning to Eq.~(\ref{formalsolfull2}) we have asymptotically, in
the leading order,
\begin{eqnarray}
\psi^{+}_{\bm{k}} ( \bm{r}) &\stackrel{r \rightarrow
\infty}{\sim}&
\phi^{(0)+}_{\bm{k}} ( \bm{r})  +
\int \frac{d {\bm k}'}{(2 \pi)^3}
\frac{T(\bm{k}',\bm{k})
\phi^{(0)-}_{\bm{k}'} ( \bm{r})
} {E-k'^2/2\mu+i\epsilon} .
\nonumber \\
\label{formalsolfull3}
\end{eqnarray}
When $r \to \infty$ the components involving bound states decrease
exponentially. Thus we have the two-body version of the asymptotic
relationship revealed in Refs.~\cite{KMS03,KMSBP03}. It states that the
leading-order asymptotic term of the scattering wave is defined by
the same (i.e., the leading-order) asymptotic term of the incident
wave. Using Eq.~(\ref{asdpwgen}) and evaluating  the integral,
taking advantage of the simple pole singularity of the integrand
at the on-shell point, we have
\begin{eqnarray}
\psi^{+}_{\bm{k}} ( \bm{r}) &\stackrel{r \rightarrow
\infty}{\sim}& e^{i\bm{k} \cdot \bm{r} + i\gamma \ln(k r -
\bm{k} \cdot \bm{r})}
\nonumber \\&&
-\frac{\mu}{2 \pi} T({k}\widehat{\bm{r}},\bm{k}) \frac{ e^{i {k}
{r}}} {r} e^{- i\gamma \ln(2 k r)}  .
\label{formalsolfull4}
\end{eqnarray}
In Eq.\ (\ref{formalsolfull4}) we dropped the modulus sign since
at this stage it is safe to do so.  By comparing Eqs.\
(\ref{formalsolfull4}) and (\ref{AB}) we conclude, as we set out
to prove, that $T(\bm{k}',\bm{k})$ introduced in
Eq.~(\ref{prior1}) is the transition matrix ($T$-matrix) which
defines the amplitude of scattering of the particles with initial
relative momentum $\bm{k}$ in the direction of $\bm{r}$:
\begin{eqnarray}
f(\widehat{\bm{k}} \cdot \widehat{\bm{r}}) &=&
-\frac{\mu}{2 \pi} T(\bm{k}',\bm{k}) ,
\label{fandT}
\end{eqnarray}
where we used a notation $\bm{k}'={k}\widehat{\bm{r}}$.  In analogy
with conventional scattering theory we call the new form for
$T(\bm{k}',\bm{k})$ as defined in Eq.~(\ref{prior1}) the {\em
prior} form of the T-matrix.

Repeating the procedure outlined above for $\psi^{-}_{\bm{k}} (
\bm{r})$ [this time we use the resolution of the total
Green's function $G$ in terms of $\psi^{+}_{\bm{k}} ( \bm{r})$] we get
\begin{eqnarray}
T^{post}(\bm{k}',\bm{k}) &=&
\big<\phi^{(0)-}_{\bm{k}'}|\loarrow{H}-E| \psi^{+}_{\bm{k}}\big> .
\label{post1}
\end{eqnarray}
In obtaining Eq.~(\ref{post1}) we also used the reciprocity
theorem, i.e.\
$T(\bm{k}',\bm{k})=T(-\bm{k},-\bm{k}')$~\cite{LL85}.
Again, in analogy with the standard theory this new form
is called the {\em post} form of the
T-matrix.

In deriving the result of Eq.~(\ref{newres}) we saw, in particular,
that
\begin{eqnarray}
  T^{prior}(\bm{k}',\bm{k}) &=& -\big<\psi^{-}_{\bm{k}'}
  |\loarrow{H}_0-\roarrow{H}_0| \phi^{(0)+}_{\bm{k}}\big> ,
  \label{ampls}
  \\
&=& - \frac{1}{2\mu} \lim_{r \rightarrow
\infty} r^2 \int d\hat{\bm{r}} \nonumber \\&& \times \left[
 \psi^{-*}_{\bm{k}'} \frac{\partial \phi^{(0)+}_{\bm{k}}}{\partial r}
- \phi^{(0)+}_{\bm{k}} \frac{\partial \psi^{-*}_{\bm{k}'} }{\partial r}
\right] . \label{prior1su}
\end{eqnarray}
In a similar way we also get from Eq.~(\ref{post1})
\begin{eqnarray}
  T^{post}(\bm{k}',\bm{k}) &=&
  \big<\phi^{(0)-}_{\bm{k}'}|\loarrow{H}_0-\roarrow{H}_0|
  \psi^{+}_{\bm{k}}\big> .  \label{post1s}
\\
&=& \frac{1}{2\mu} \lim_{r \rightarrow
\infty} r^2 \int d\hat{\bm{r}} \nonumber \\&& \times \left[
 \phi^{(0)-*}_{\bm{k}'} \frac{\partial \psi^{+}_{\bm{k}}}{\partial r}
-\psi^{+}_{\bm{k}} \frac{\partial \phi^{(0)-*}_{\bm{k}'} }{\partial r}
\right] .
\label{post1su}
\end{eqnarray}
Thus the scattering  T-matrix conventionally given as volume
integrals can be written equivalently in surface-integral forms.
We emphasize that in these
forms the T-matrix depends only on the asymptotic behavior of the
participating functions. Therefore, knowledge of the scattering wave
function over the entire space is not required.
In addition, the surface--integral forms are readily expanded in
partial waves leading to a simple result containing only the limiting
procedure.
Therefore, these forms are particularly suitable for practical
calculations. It is also interesting to note the close resemblance
of these forms to the representation of the number of scattered
particles crossing the surface element $d\hat{\bm{r}}$ per unit
time at large distance~$r$. In that sense, the surface--integral
forms further reveal the scattering amplitude as the amplitude of
the probability flux of particles scattered in
direction~$\hat{\bm{r}}$.

We note that in an operator form Eq.~(\ref{formalsolfulshort})
can be written as
\begin{eqnarray}
\psi^{\pm}_{\bm{k}} ( \bm{r}) &=&
[ 1+G(E \pm i\epsilon)(\roarrow{H} -E) ]\phi^{\pm}_{\bm{k}} ( \bm{r}) ,
\label{formali0}
\end{eqnarray}
where $G(E\pm i\epsilon)=(E\pm i\epsilon-H)^{-1}$ is the Green's
operator associated with the Green's function $G(\bm{r},
\bm{r}';E\pm i\epsilon)$.
Therefore, we can introduce new generalized wave operators according to
\begin{eqnarray}
\Omega^{\pm} &=&
[ 1 +G(E\pm i\epsilon)(\roarrow{H}
-E) ] . \label{waveoper}
\end{eqnarray}
In the next section we turn to a further investigation of these
generalized wave operators.

\section{Consistency with conventional results}

Our aim in this section is to show the results given above are
consistent with conventional potential scattering theory for
short-range interactions. The existing formulation of scattering
theory relies on the condition that interaction $V(r)$ decreases
faster than the Coulomb interaction when $r \to \infty$
[$\gamma=0$ in Eqs.~(\ref{casseries}) and (\ref{cas})], so that
\begin{eqnarray}
\phi^{\pm}_{\bm{k}} ( \bm{r}) =  \phi^{(0)\pm}_{\bm{k}} ( \bm{r}) \to  \phi^{(0)}_{\bm{k}} ( \bm{r}) =e^{i\bm{k} \cdot \bm{r}} . 
\end{eqnarray}
The initial unscattered wave function satisfies the Helmholtz equation
\begin{eqnarray}
  (E -H_0) \phi^{(0)}_{\bm{k}} ( \bm{r}) =0 . \label{HE}
\end{eqnarray}
Then
\begin{eqnarray}
  (H -E) \phi^{(0)}_{\bm{k}} ( \bm{r}) = V \phi^{(0)}_{\bm{k}}
  ( \bm{r}) , \label{HE2}
\end{eqnarray}
therefore, Eq.~(\ref{formalsolfulshort}) takes the form
\begin{eqnarray}
  \psi^{\pm}_{\bm{k}} ( \bm{r}) &=& \phi^{(0)}_{\bm{k}} ( \bm{r})+
  \int d\bm{r}'
  G(\bm{r}, \bm{r}';E \pm i\epsilon)
\nonumber \\
&&\times  (\roarrow{H} -E) \phi^{(0)}_{\bm{k}} (
  \bm{r}') 
\nonumber \\
&=& \phi^{(0)}_{\bm{k}} ( \bm{r})+
  \int d\bm{r}'
  G(\bm{r}, \bm{r}';E \pm i\epsilon)
  V \phi^{(0)}_{\bm{k}} (
  \bm{r}') .
\nonumber \\
\label{formalsolold}
\end{eqnarray}
This is the formal solution which is obtained in conventional
scattering theory.  In the light of Eq.~(\ref{HE2}),
Eqs.~(\ref{prior1}) and~(\ref{post1}) reduce to
\begin{eqnarray}
T^{prior}(\bm{k}',\bm{k}) &=& \big<\psi^{-}_{\bm{k}'} |V|
\phi^{(0)}_{\bm{k}}\big> , \label{amplshort}
\\
T^{post}(\bm{k}',\bm{k}) &=&  \big<\phi^{(0)}_{\bm{k}'}|V|
\psi^{+}_{\bm{k}}\big> ,  \label{post1short}
\end{eqnarray}
in agreement with the standard definitions of the T-matrix.
Moreover, for the same reason the generalized wave operators
$
\Omega^{\pm}$ introduced above reduce to the usual M\"oller (M)
ones:
\begin{eqnarray}
\Omega^{\pm}_M &=&
[ 1 +G(E\pm i\epsilon) V ]
. \label{moller}
\end{eqnarray}
Obviously, when interaction $V$ has a tail which does not
disappear at infinity, the Helmholtz equation~(\ref{HE}) for
$\phi^{(0)+}_{\bm{k}} $, and consequently Eq.~(\ref{HE2}), are not
satisfied. As a result
Eqs.~(\ref{formalsolold}),~(\ref{amplshort})
and~(\ref{post1short}) are incorrect and Eq.\ (\ref{moller}) is
not valid for Coulomb--like potentials.

On the other hand, when the interaction is purely Coulomb, we can
proceed further with analytical methods. Then we have
\begin{eqnarray}
(H-E) \phi^{(0)\pm}_{\bm{k}} ( \bm{r}) =\frac{\gamma^2 k}{\mu r(kr \mp \bm{k} \cdot
\bm{r})}  \phi^{(0)\pm}_{\bm{k}} (\bm{r}) .
\label{bra}
\end{eqnarray}
Therefore, Eqs.\ (\ref{prior1}) and (\ref{post1}) transform to
\begin{eqnarray}
T^{prior}(\bm{k}',\bm{k}) &=&
\big<\psi^{-}_{\bm{k}'} |
\frac{\gamma^2 k}{\mu r(kr-\bm{k} \cdot \bm{r})} |
\phi^{(0)+}_{\bm{k}}\big> ,
\label{culampprior}
\\
T^{post}(\bm{k}',\bm{k}) &=&
\big<\phi^{(0)-}_{\bm{k}'} |
\frac{\gamma^2 k}{\mu r(kr+\bm{k} \cdot \bm{r})} | \psi^{+}_{\bm{k}}\big>
\label{culamppost} .
\end{eqnarray}
Here $\psi^{\pm}_{\bm{k}}$ are known and given by
\begin{eqnarray}
\psi^{\pm}_{\bm{k}} ( \bm{r}) &=& e^{i\bm{k} \cdot \bm{r}} e^{-\pi
\gamma/2} \Gamma (1 \pm i\gamma)
\nonumber \\&&\times
_1F_1(\mp i\gamma,1,\pm i kr-i\bm{k} \cdot \bm{r}) ,
\label{cwf}
\end{eqnarray}
with
$_1F_1$ being the usual confluent hypergeometric function.
At the same time Eq. (\ref{formalsol0}) transforms to
\begin{eqnarray}
\psi^{(0)\pm}_{\bm{k}} ( \bm{r}) &=& \phi^{(0)\pm}_{\bm{k}} (
\bm{r})
+
\int d\bm{r}'
G(\bm{r}, \bm{r}';E\pm i\epsilon) \nonumber \\&& \times
\frac{\gamma^2 k}{\mu r'(kr' \mp \bm{k} \cdot \bm{r}')}
\phi^{(0)\pm}_{\bm{k}} ( \bm{r}') . \label{BaMacorrect}
\end{eqnarray}
This is to be compared with the result obtained in Eq.~(16) of
Ref.~\cite{BM89} for the total scattering wave function~$\psi^{\pm}_{\bm{k}}$:
\begin{eqnarray}
\psi^{\pm}_{\bm{k}} ( \bm{r}) &=& \phi^{(0)\pm}_{\bm{k}} (
\bm{r})
+
\int d\bm{r}'
G(\bm{r}, \bm{r}';E\pm i\epsilon) \nonumber \\&& \times
\frac{\gamma^2 k}{\mu r'(kr' \mp \bm{k} \cdot \bm{r}')}
\phi^{(0)\pm}_{\bm{k}} ( \bm{r}') . \label{BaMa}
\end{eqnarray}
If Eq.\ (\ref{BaMa}) were true it would mean that
\begin{eqnarray}
\int d\bm{r}'
G(
\bm{r}, \bm{r}';
E \pm i\epsilon)
(E-\roarrow{H}) \left[ \psi^{\pm}_{\bm{k}} ( \bm{r}')
-\phi^{(0)\pm}_{\bm{k}} ( \bm{r}')  \right]
\nonumber \\
=
\psi^{\pm}_{\bm{k}} ( \bm{r})
-\phi^{(0)\pm}_{\bm{k}} ( \bm{r}) ,
\label{wrong}
\end{eqnarray}
which is, however, not correct (see Eq.~(\ref{neweq1}) and discussion
following it).
Based on Eq.~(\ref{BaMa}) Barrachina and Macek also arrived at
Eq.\ (\ref{culamppost}), nevertheless, since the underlying
equation was not correct this result is not justified
in~\cite{BM89}.  On a positive note, the matrix elements in Eqs.
(\ref{culampprior}) and (\ref{culamppost}) (to be more precise,
the complex conjugate of the former) have been evaluated in
Ref.~\cite{BM89} in closed form. We have checked and confirm their
result, namely
\begin{eqnarray}
\lefteqn{
\big<\psi^{-}_{\bm{k}'} |
\frac{\gamma^2 k}{\mu r(kr-\bm{k} \cdot \bm{r})}
| \phi^{(0)+}_{\bm{k}}\big>
}
\nonumber \\
&\equiv&
\big<\phi^{(0)-}_{\bm{k}'} |
\frac{\gamma^2 k}{\mu r(kr+\bm{k} \cdot
\bm{r})} | \psi^{+}_{\bm{k}}\big>
\nonumber \\&=&
\frac{4 \pi z_1 z_2}{|\bm{k}'-\bm{k}|^2}
\frac{\Gamma(1+i\gamma)}{\Gamma(1-i\gamma)} \left[\frac{4
k^2}{|\bm{k}'-\bm{k}|^2}\right]^{i\gamma} ,
\end{eqnarray}
which is the well-known full on-shell Coulomb T-matrix.
This gives additional support for the new definitions of the
T-matrix.

Finally, in the pure Coulomb case
the generalized wave operators $\Omega^{\pm}$ introduced earlier
reduce to the wave operators obtained by Mulherin and Zinnes
(MZ)~\cite{MZ70}:
\begin{eqnarray}
\Omega^{\pm}_{MZ} &=&
\left[ 1 +G(E\pm i\epsilon)\frac{\gamma^2 k}{\mu r(kr \mp \bm{k} \cdot
\bm{r})} \right]
, \label{mzwaveoper}
\end{eqnarray}
provided $
\Omega^{\pm}$ are applied to the first term of the incident wave
$\phi^{(0)\pm}_{\bm{k}}$. This clearly shows that the MZ operators are
approximations to the corresponding full wave operators. Obviously,
this approximation makes a sense only for asymptotically large
distances where $\phi^{(0)\pm}_{\bm{k}}$ becomes the dominant
(leading-order) term. Elsewhere, the MZ operator cannot be relied
upon. This finding may also explain formal problems associated with
the MZ approach~\cite{Dollard75}.

\section{Conclusion}

In this paper we have presented a generalization of potential
scattering theory which is valid for arbitrary interactions including
potentials with the long-range Coulomb tail. We obtained a new formal
solution to the Schr\"odinger equation satisfying the boundary
conditions imposed by the long-range nature of the Coulomb
interaction.  We introduced new general definitions for the scattering
amplitude and wave operators.  We showed that when the interaction
potential is short-ranged the generalized definitions of the
scattering amplitude and wave operators transform to the conventional
ones used in standard scattering theory. A distinctly satisfying
feature of the new forms for the scattering amplitude is that they do
not contain divergent factors and directly give the physical on-shell
scattering amplitude even when the interaction potential is
long-ranged. Moreover, the generalized wave operators are also the
same for arbitrary potentials including the long-range interactions.
Therefore, no modification of the theory based on renormalization is
required.  The results of the present work close the gap between the
two different formulations of potential scattering theory for short-
and long-range interactions.

In conclusion we make the following comments. In the literature
devoted to the Coulomb scattering it is customary to separate
$\psi^{\pm}_{\bm{k}}$ into the `incident' and the `scattered' parts
according to
\begin{eqnarray}
  \psi^{\pm}_{\bm{k}} ( \bm{r})&=&
  \tilde \phi^{\pm}_{\bm{k}} ( \bm{r})+
  \tilde \chi^{\pm}_{\bm{k}}( \bm{r}) , \label{splitold}
\end{eqnarray}
where the incident wave is taken as~\cite{H85}
\begin{eqnarray}
 \tilde \phi^{\pm}_{\bm{k}} ( \bm{r}) &=&
  e^{i\bm{k} \cdot \bm{r}} e^{\pi
    \gamma/2} \, U(\mp i\gamma,1, \pm i kr-i\bm{k} \cdot \bm{r}) . \label{casfull}
\end{eqnarray}
Here $U$ is the confluent hypergeometric function of the second kind.
The idea is driven by the fact that the confluent hypergeometric
function in the regular Coulomb function [see Eq.~(\ref{cwf})] can be
written as a sum of two irregular confluent hypergeometric functions
of the second kind and that functions $\tilde \phi^{\pm}_{\bm{k}} (
\bm{r})$ and $\tilde \chi^{\pm}_{\bm{k}}( \bm{r})$ satisfy the first
and the second parts of asymptotic conditions~(\ref{AB}) or
(\ref{AB-}), respectively.  However, (for the pure Coulomb
interaction) $\tilde \phi^{\pm}_{\bm{k}} ( \bm{r})$ alone is a
solution to the original Schr\"odinger equation, i.e.:
\begin{eqnarray}
  (E -H) \tilde \phi^{\pm}_{\bm{k}} ( \bm{r}) =0 . \label{SEphi}
\end{eqnarray}
Consequently, the scattered wave $\tilde \chi^{\pm}_{\bm{k}}$ is a
solution as well. Thus, as a result of separation (\ref{splitold})
Eq.~(\ref{SE}) splits into two equations making it impossible to
single out uniquely the important surface-integral components in
the full solution. Therefore representation (\ref{splitold}) is
not a satisfactory starting point.  It also leads to other
anomalies associated with the Coulomb problem.  In particular,
using Eq.~(\ref{SEphi}) one can demonstrate that the Coulomb wave
function is a solution to a {\em homogeneous} Lippmann-Schwinger
equation~\cite{W67}.  On the other hand the splitting
(\ref{split}) used in this work represents the logical fact that
the `unscattered' incident wave is coming from infinity and should
be taken in a form valid at asymptotically large distances. 

We conclude by offering a few comments on the possible usefuleness of
our methods for three-body systems with long-range interactions.
Rigorous scattering theory for a system of three particles valid for
short-range potentials was given by
Faddeev~\cite{Faddeev60Ru,Faddeev63}.  For the charged
particles with the long-range Coulomb interaction the theory has faced
apparently insurmountable difficulties.  The problem is that the
Faddeev equations are not compact in the presence of Coulomb
interactions. In other words these equations cannot be solved using
standard numerical procedures.
Some progress has been made in dealing with aspects of the
three-body problem with Coulomb-like potentials. In particular, a
renormalization method based on screening~\cite{D68,Taylor75} has
been implemented successfully for the case when two particles are
charged~\cite{ASZ78,AS96}. The method has also been extended to
two-fragment reactions in a system of three charged
particles~\cite{AS80,AS96}. Though Dollard's time-dependent
approach~\cite{D64,Dollard71} is formally valid for arbitrary
multichannel collisions including three-body problem, no practical
time-independent renormalization method exists that is valid for a
system of three charged particles above the breakup threshold. The
problem is that above the threshold the Coulomb three-body system
possesses essentially different types of singularities and the
two-particle renormalization procedures are not sufficient to
guarantee compactness of the
equations~\cite{V70Ru,Mer77Ru,Mer80}. Thus, on one
hand there are no integral equations yet known for collisions of
more than two charged particles that are satisfactory above the
breakup threshold~\cite{H85}, and on the other hand, there is also
neither theoretical proof nor practical evidence that
renormalization approach can be applied to the Faddeev equations
for the genuine three-body Coulomb problem. This is a rather
disturbing situation especially for the atomic three-body problem
where all three particles are charged. Therefore, generally
speaking, it would be useful to formulate three-body scattering
theory in a manner that does not require renormalization so that
the aforementioned modifications are, in a certain sense,
unnecessary. We are confident that the method proposed here for
solving the two-body problem that was free of the usual Coulomb
anomalies can be profitably applied to the proper formulation of the
three-body rearrangement theory for long-range interactions.

\acknowledgments

The work was supported by the Australian Research Council, U.S. DOE
Grant DE-FG03-93ER40773 and NSF Grant PHY-0140343.


\begin{thebibliography}{38}
\bibitem{endnote}{Here and in what follows we omit the symbol
limit as ${\epsilon \to 0}$.  This should not lead to a
    misunderstanding.}
\bibitem{endnote2}{In the presence of the Coulomb potential this
    expansion diverges as a function. However, it has been shown to
    converge as a distribution~\cite{Taylor74}. Therefore, the purely
    formal partial-wave manipulations used in scattering theory does
    make sense also for the Coulomb case.}
\end{thebibliography}

\end{document}